\documentclass[color]{aastex62}
\usepackage{tabularx}
\usepackage{epstopdf}
\usepackage{ulem}
\received{July 23, 2018}
\revised{September 11, 2018}
\accepted{September 13, 2018}
\submitjournal{ApJ}

\shorttitle{Monte Carlo modeling of the Compton shoulder of Circinus}
\shortauthors{Hikitani et al.}

\begin{document}

\title{Compton Shoulder Diagnostics in Active Galactic Nuclei for 
Probing the Metalicity
of the Obscuring Compton-Thick Tori}

\correspondingauthor{M. Ohno}
\email{ohno@astro.hiroshima-u.ac.jp}

\author{Masaya Hikitani}
\affiliation{School of Science, Hiroshima University, 1-3-1, Kagamiyama, Higashi-Hiroshima, Hiroshima 739-8526, Japan}

\author{Masanori Ohno}
\affiliation{School of Science, Hiroshima University, 1-3-1, Kagamiyama, Higashi-Hiroshima, Hiroshima 739-8526, Japan}

\author{Yasushi Fukazawa}
\affiliation{School of Science, Hiroshima University, 1-3-1, Kagamiyama, Higashi-Hiroshima, Hiroshima 739-8526, Japan}

\author{Toshihiro Kawaguchi}
\affiliation{Faculty of Economics, Management and Information Science, Onomichi City University, Onomichi, Hiroshima 722-8506, Japan}


\author{Hirokazu Odaka}
\affiliation{Department of Physics, The University of Tokyo, 7-3-1 Hongo, Bunkyo-ku, Tokyo 113-0033, Japan}



\begin{abstract}

We analyzed the spectral shape
 of the Compton shoulder around the neutral Fe-K$_\alpha$ line of the Compton-thick type II Seyfert nucleus of the Circinus galaxy. 
The characteristics of this Compton shoulder with respect to the reflected continuum and Fe-K$_\alpha$ line core intensity are a powerful diagnostics tool for analyzing the structure of the molecular tori, which obscure the central engine.
We applied our 
Monte-Carlo-based X-ray reflection spectral model to the {\it Chandra} High Energy Transmission Grating data and successfully constrained the various spectral parameters 
independently, using only the spectral data only around the Fe-K$_\alpha$ emission line. The obtained column density and inclination angle are consistent with the previous observations and the Compton-thick type II Seyfert picture. In addition, we determined the metal abundance of the molecular torus for the case of the smooth and clumpy torus to be 1.75$^{+0.19}_{-0.17}$ and 1.74$\pm$0.16 solar abundance, respectively. Such slightly over-solar abundance 
can be useful information for discussing the star formation rate in the molecular tori of active galactic nuclei.

\end{abstract}


\section{Introduction} \label{sec:intro}

The large amount of luminosity of 
active galactic nuclei (AGNs) is sustained 
by gas accretion onto
supermassive black holes (BHs) at the center of each galaxy.
The accretion disk and the BH are surrounded by an optically 
thick, dusty torus
 (e.g., \cite{1984ApJ...282..427T}; 
 \cite{1985ApJ...297..621A}). 
%
A large geometrical thickness of the torus 
(\cite{1993ARA&A..31..473A}; \cite{1991MNRAS.252..586L})
 indicates
that numerous dusty clumps, 
rather than a smooth mixture of gas and dust, 
constitute the torus with a large clump-to-clump velocity 
dispersion 
(\cite{1988ApJ...329..702K}). 

The torus is illuminated by the central accretion disk 
(e.g., 
\cite{2004MNRAS.350.1049G}; 
\cite{2004ApJ...612L.113S}). 
Since the disk illumination is weak toward its mid-plane, 
the innermost edge of the torus can connect 
with the outermost edge of the accretion disk continuously 
(\cite{2010ApJ...724L.183K}). 
Specifically, the 
torus potentially acts as a role of a 
reservoir 
for the gas inflow to the accretion disk, with mutual influences.
Although 
various clumpy torus models for the 
infrared emission of AGNs 
have been developed 
(e.g., \cite{2002ApJ...570L...9N,2008ApJ...685..147N}; 
\cite{2006A&A...452..459H}), 
the observed near-infrared emission is not explained by these models in many cases 
(e.g., \cite{2012MNRAS.420..526M}; 
\cite{2014ApJ...785..154L}). 
Accordingly, fitting for the observed mid-infrared emission is not 
enough for solving the degeneracy among various parameters 
(the clump size, 
the metalicity 
and the viewing angle etc.).

The Compton shoulder (recognized as a residual
at the low-energy side of a narrow Fe-K$_\alpha$ line) has been clearly seen in Galactic BH binaries 
(\cite{2003ApJ...597L..37W,2010ApJ...715..947T}).
For a handful
AGNs, detection of the Compton shoulder has been reported in  
grating data 
(e.g., \cite{2011ApJ...738..147S}; 
\cite{2002ApJ...574..643K}),  
as well as in conventional CCD data
(e.g., \cite{2003MNRAS.343L...1M,2002A&A...396..793B,1997MNRAS.289..443I}). 
An additional narrow line 
adjacent to Fe-K$_\alpha$ is often used to model/mimic 
the Compton shoulder 
(e.g., \cite{2004A&A...414..155M}). 
The flux ratio of the Compton shoulder with respect to the line core depends 
on both the column density of the reflection material $N_{\rm H}$ 
and the Fe abundance (\cite{2002MNRAS.337..147M,2016ApJ...818..164F}).
For example, although the metalicity of the Circinus galaxy is estimated 
to be 1.2--1.7 solar via the depth of the Fe K edge 
(\cite{2003MNRAS.343L...1M}), 
the depth, in principle, 
also depends on $N_{\rm H}$ 
(\cite{2015MNRAS.454..973Y}). 
The Fe-K$_\alpha$ flux and the continuum shape also depend on these parameters, 
and hence X-ray spectroscopic data of binaries and AGNs 
around the Fe-K$_\alpha$ line enables us to constrain various parameters 
(\cite{2016ApJ...818..164F,2016MNRAS.462.2366O}). 

In this paper, we applied a developed X-ray reflection spectral model for the two types of density structures (smooth and clumpy) of the 
torus of the AGN to constrain these parameters independently, focusing on the 
structure of the Compton shoulder. 
In the next section, the data we used is briefly introduced. 
Then, we present the spectral modeling (\S 3) and the results (\S 4).
Finally, we carry out discussions, followed by the conclusion 
of this study. 

\section{Observations and Data Reduction} \label{sec:obs}

\cite{2011ApJ...738..147S} reported the analysis result of 10 Compton-thick type II Seyfert galaxies observed by
the {\it Chandra} High Energy Transmission Grating ({\it HETG}) and their detailed analysis for the Fe-K$_\alpha$ line width constrained the
line-emitting region. Because of the excellent energy resolution of the {\it Chandra HETG}, some of their 
samples show a hint of the spectral structure due to the Compton shoulder in the foot region of the Fe-K$_\alpha$ line emission, 
such as the galaxies NGC1068, Centaurus A, NGC4388, and Circinus galaxy. The Fe-K$_\alpha$ line intensity of Circinus
galaxy is highest among the samples of \cite{2011ApJ...738..147S} and the structure of the Compton shoulder was the most distinctive.
Therefore, we selected Circinus galaxy for our analysis. 

The Circinus galaxy was observed by the {\it Chandra HETG} four times from June 2000 to
November 2004, with a net total exposure of about  185 ks. Table \ref{tab:obs_summary} gives a summary of these observations. We utilized the archival {\it Chandra} data and reduced the data with the software package {\it Chandra Interactive Analysis of Observations (CIAO)} v4.10 with the 
calibration data base (CALDB) version of 4.7.3, followed by standard analysis procedures described in the {\it CIAO} analysis thread for the HETG/ACIS-S Grating Spectra. The {\it chandra\_repro} task
has been applied for the standard data distribution to extract grating spectra and detector responses. Although higher orders of the grating data provide a higher energy resolution, we used only the first orders of the grating data because only the first orders of the data allow investigation of the feature of the Compton shoulder for its better photon statistics. We combined the positive and negative arms with 
the {\it combine\_grating\_spectra} script provided by {\it CIAO}. The background spectrum was extracted from the original combined data format by using 
the {\it tg\_bkg} script so that we could deal with the {\it Chandra HETG} spectra using the {\it XSPEC}. 
Finally, we combined all spectra and detector responses of four observations into one file with the {\it mathpha} and
{\it addrmf} tools in the standard {\it FTOOLS}. The reduced spectrum was analyzed by using the spectral fitting tool {\it XSPEC} (\cite{1996ASPC..101...17A}).
We utilized the {\it C}-statistics for the minimization without any energy-binning procedures. All the quoted errors in the following sections were 90\% confidence level. 
The spectral fitting was employed in the 4 $-$ 8 keV energy band to avoid any 
contaminations of the complicated spectral features, which are often reported for the type-II Seyferts in the soft X-ray energy band (\cite{2006A&A...448..499B}).

\begin{table*}[htbp]
\begin{center}
\caption{Observation Summary of the {\it Chandra HETG} for the Circinus Galaxy}
\label{tab:obs_summary}
\begin{tabularx}{15cm}{l|l|l|l|l|l}
\hline
\hline
Source & Obs Position [R.A.  Dec.] & Redshift & ObsID & ObsDate & Exposure [s]\\
\hline
Circinus galaxy & 14 13 10.20, -65 20 20.6 & 0.001448 &  374 & 2000-06-15 22:01:09 & 7260 \\
			& 14 13 10.20, -65 20 20.6 & &  62877 & 2000-06-16 00:38:28 & 61400 \\
			& 14 13 10.20, -65 20 20.8 & &  4770 & 2004-06-02 12:40:42 & 56100 \\
			& 14 13 10.20, -65 20 20.8 & &  4771 & 2004-11-28 18:26:32 & 60180 \\
\hline
\end{tabularx}
\end{center}
\end{table*}

 The {\it Chandra HETG} grating spectroscopy is performed by analyzing the dispersion image, which is accumulated by the X-ray telescope. By extracting dispersion spectra following the standard {\it chandra\_repro} processes, the extracted zeroth-order, non-dispersed image extends up to at least the 5" scale, which is larger than the PSF of the {\it Chandra HRMA}. Therefore, our analyzed spectrum could include not only the central torus region but also a more extended outer region. The central torus region, whose radius of $\sim$\,30\,pc ($<(1-1.5)"$), based on the ALMA observation (Izumi et al. in prep), is considered the main contributor of   the Fe-K$_\alpha$ line for Circinus (\cite{2013MNRAS.436.2500M})
 , and thus we fit the extracted X-ray data with our torus model.

\section{Spectral Modeling and Analysis} \label{sec:analysis}

\begin{figure}[tbp]
\begin{center}
\rotatebox{-0}{\resizebox{10cm}{!}{\includegraphics{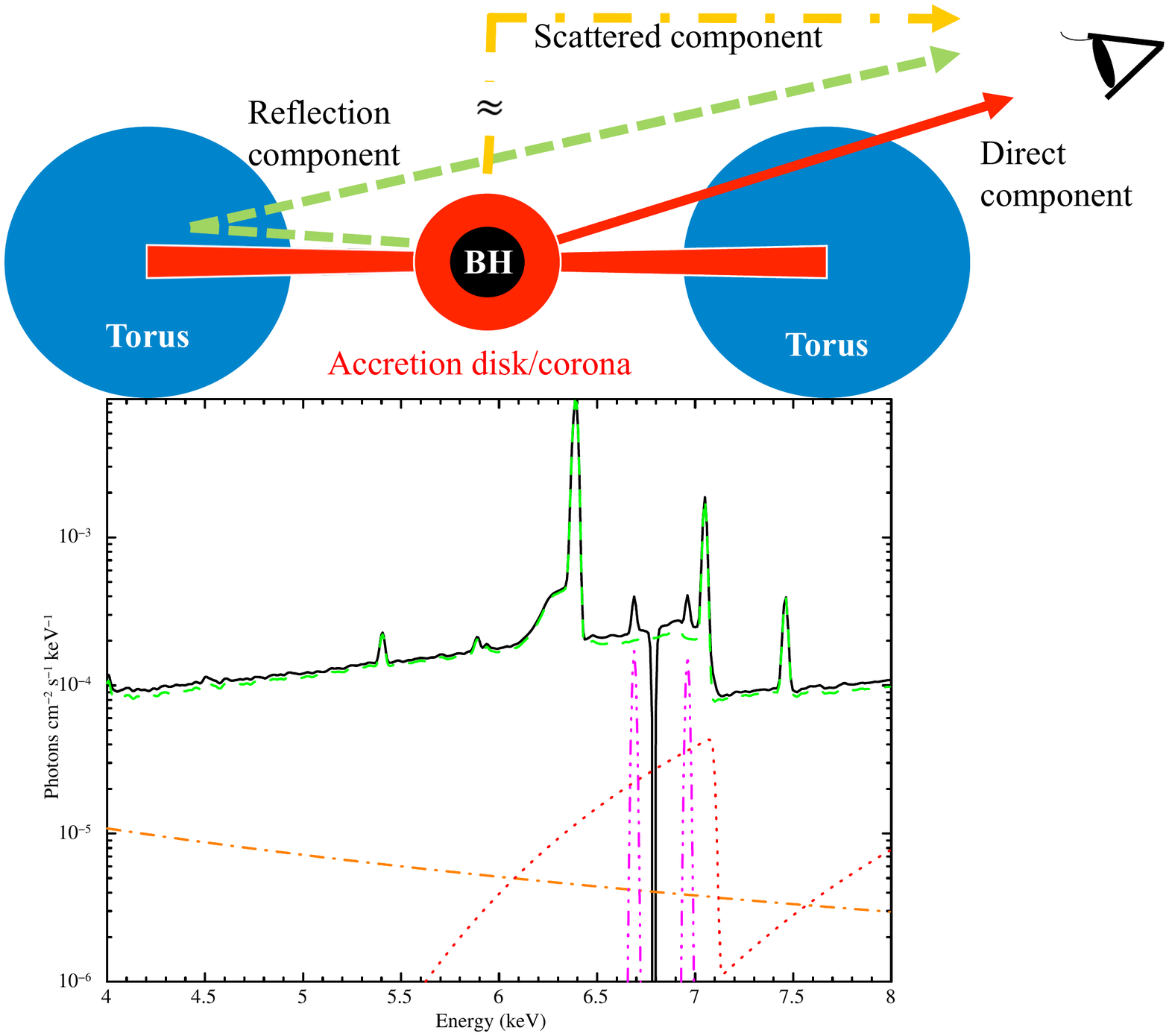}}}
\caption{Schematic picture of our baseline model. The solid, dotted, and dot-dashed arrow lines represent the direct, reflection, and
scattered component, respectively, which are mainly introduced in this analysis. An example of the simulated spectrum is also shown, 
where the contribution of each component to the overall spectrum can be seen.}
\label{fig:our_model}
\end{center}
\end{figure}

The observed X-ray emission from obscured AGNs consists of several continuum 
components and complicated emission line profiles. A Comptonized emission by the hot corona 
around the central black hole can be observed as the simple power-law shape up to the hard X-ray regime. 
However such a "direct" component cannot be observed below 10 keV for the Compton-thick object, 
due to the strong absorption by the surrounding torus. Instead, various reflected and scattered emissions
are observed in this energy range. The reflection component from the Compton-thick tori consists of a
 Compton scattered hard spectrum with various line emissions, including a prominent Fe-K$\alpha$ line
 and its Compton shoulder. For analyzing the characteristics of this Compton shoulder from the
 reflection component, which is the main goal of this study, an accurate modeling of the reflected and scattered continuum
 is essential. We consider other additional spectral components: the scattered continuum from a distant region such 
 as the narrow line region and various ionized lines that are not included in our current reflection model.
 Figure \ref{fig:our_model} shows a schematic picture of our baseline model used for this analysis.
 The details of the modeling for each component are given below.
 
\subsection{Reflection Component from Compton Thick Torus} \label{sec:furui_model}

\cite{2011ApJ...740..103O} developed a Monte Carlo simulation framework to calculate the radiation transfer in the astronomical object
(MONACO) and we have already succeeded in reproducing the reflection component from the AGN by applying this framework (\cite{2016ApJ...818..164F}).
This model has many parameters, including metal abundance, absorption column density, density distribution inside the torus (smooth or clumpy), etc.
\cite{2016ApJ...818..164F} found that these parameters influence the
strength and the shape of the Compton shoulder, indicating
that the Compton shoulder can solve the degeneracy among
various parameters by using data with high enough energy resolution.
\cite{2016MNRAS.462.2366O} 
demonstrated  that a degeneracy between
the metalicity and the column density of a reflection
material (for a slab or a spherical geometries) is solved by utilizing the
{\it Chandra HETG} data of 
a high-mass X-ray binary GX~301-2.
We apply the torus model to the Compton shoulder 
of the AGN to comprehend the extent to which the degeneracies of these parameters 
are solved.
In this analysis, we free three parameters for our 
torus model; hydrogen 
column density at the torus mid-plane $N_{\rm H}$, metal abundance (MA),
and the inclination angle for the observer ($\theta_{ia}$) as well as the normalization. 
We changed the abundance of all metals heavier than beryllium, keeping the same ratio to the solar abundance table from \cite{1989GeCoA..53..197A}. 
The important parameter of this analysis is the normalization ratio between Fe-K emission line and its Compton shoulder. Therefore, we can say that if we change only iron abundance, our results would not be changed so much. 
While, We fixed other several 
parameters; the opening angle of the torus, the innermost radius of the torus, and the turbulent velocity of the torus element are set to 60 degree,
2 $\times$ 10$^6$ cm, and 0 km s$^{-1}$, respectively. 
In 
the clumpy model,
the volume-filling factor of the clump and the 
scale factor of each clump (the ratio of radii between 
each clump and the torus) are set to 0.05 and 0.005, respectively. 
According to \cite{2016ApJ...818..164F}, the parameters related to the clumpy torus (the volume-filling factor and scale factor of clumps) do not affect to the shape and ratio with respect to the continuum of the Compton shoulder which are important features to determine three parameters that we focused on here, and other parameters related to the torus 
geometry could affect to the normalization of the reflection but would not change features around the Compton shoulder.

In addition, we 
set the absorption column density of the direct (transmitted) component (\S \ref{sec:direct}) 
to $\sqrt{(1-4{\rm cos}\theta_{ia})}$ times $N_{\rm H}$, 
 in order to 
take into account the fact that our simulation model assumes the donut-shaped material distribution of the torus.
The spectral shape of emission lines, including Fe-K$_\alpha$, and their Compton shoulder are
calculated based on the Monte Carlo simulation. We examined both the "smooth" and "clumpy" torus models  to constrain the density structure of the 
torus with the characteristics of the Compton shoulder. 
Then, we discuss how and to what extent the derived parameters
differ from each other.

\subsection{Direct Continuum from Nucleus} \label{sec:direct}
We introduce the direct emission component, 
albeit little contribution for Compton-thick objects, 
since it affects to
the estimation of the intensity of the scattered continuum component as described below. 
The spectral model of this component is the
simple power-law model with a fixed photon index value of 1.9, which is a typical value for the type-I AGNs whose direct component is observed. 
It is difficult to constrain the normalization of
this direct component by the {\it Chandra HETG} data due to the limited energy band. 
In the spectral fitting, the normalization is scaled so that the 
Compton reflection component (\S \ref{sec:furui_model}) and this direct component 
have similar number counts around 7.0 keV for 
$N_{\rm H} = 10^{24}$\,cm$^{-2}$, which is found in 
our Monte Carlo simulation for an 
inclination angle cos$\theta_{ia} \sim$ 0.3 and 
$N_{\rm H} \sim$ 10$^{24}$ cm$^{-2}$.  

\subsection{Scattered Continuum from Distant Region} \label{sec:reflection_NLR}

Sometimes, a power-law component is required in addition to the reflection component, even for the Compton-thick object at a softer energy band. 
The spectrum of this additional power-law, with the photon index identical
to the primary power-law (1.9 in this analysis),
is much softer than the reflection. 
This additional component is believed to come from the scattering
of the direct component by less dense 
material located at the distant region compared with the Compton-thick torus, 
for instance the narrow line region 
(\cite{2000ApJ...542..175A,2008PASJ...60S.293A,2009A&A...496..653M}). 
The fitted structure of the Compton shoulder should be sensitive to the reflected continuum, 
 and fitting for the 
 reflected continuum 
is also affected by such an additional power-law component. 
Therefore, we added this component to our model. 
We set the constraint on the normalization parameter as the ratio with respect to the
direct component (before attenuation by the torus). 
This flux ratio is estimated 
to be between 0.001--0.05
(\cite{1991ApJ...375...78B},\cite{1993RMxAA..27...73B},\cite{2004ApJ...617..930M}). 
We found that the value of 0.001 gives the best fit to our data.
Thus we apply this ratio in the following analysis.

\subsection{Ionized Emission Lines} \label{sec:EL}

In the X-ray spectrum from the AGN, not only a neutral Fe-K$_\alpha$ emission but also other emission lines from highly ionized materials are often observed 
(\cite{1997MNRAS.289..443I,2004A&A...414..155M}). 
We found that  the {\it Chandra HETG} spectrum of the Circinus galaxy also show complicated line structures around the 6.5 $\sim$ 7.0 keV energy band. Firstly, we added multiple Gaussian functions to
represent all the emission and absorption features, 
but we found that it was difficult identify the origin of most of the fitted lines and it did not affect 
the structure of the Compton shoulder. Although the $\Delta \chi^2$ value is not negligible ($\Delta \chi^2 \sim 50$ for the 580 to 575 change of the degree of freedom), we  just left two most prominent and easily understandable emission lines and one absorption line, which come from He- and H-like iron (6.70 keV and 6.97 keV for emission lines and around 6.79 keV for the absorption line), respectively. 

We also considered a galactic absorption and line broadening due to the random motion of the 
emitting/reflecting gas
by the Gaussian convolution with the {\it gsmooth} model in {\it XSPEC}.
Since our reflection model calculated the spectrum in only a source frame, we introduce the redshift effect by using the {\it zashift} model. Therefore, our baseline model consists of the following formula; $ gsmooth \times zwabs ( PL_{scat} + zwabs\times PL_{direct} +zashift \times Reflection + emission\_ lines)$.

\section{Results} \label{sec:result}

\begin{figure*}[htbp]
\begin{center}
\begin{minipage}{8cm}
\rotatebox{-90}{\resizebox{6cm}{!}{\includegraphics{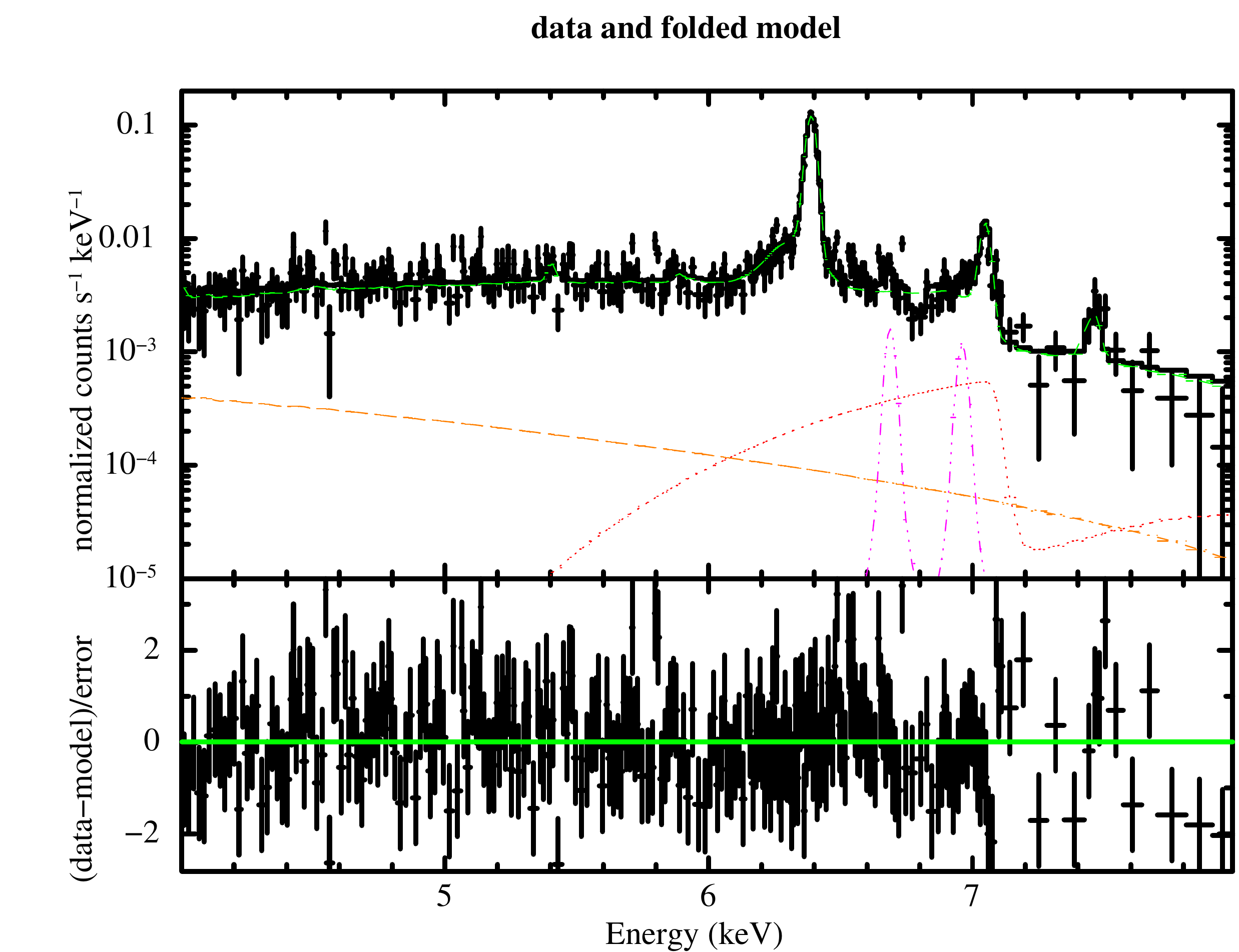}}}
\end{minipage}
\begin{minipage}{8cm}
\rotatebox{-90}{\resizebox{6cm}{!}{\includegraphics{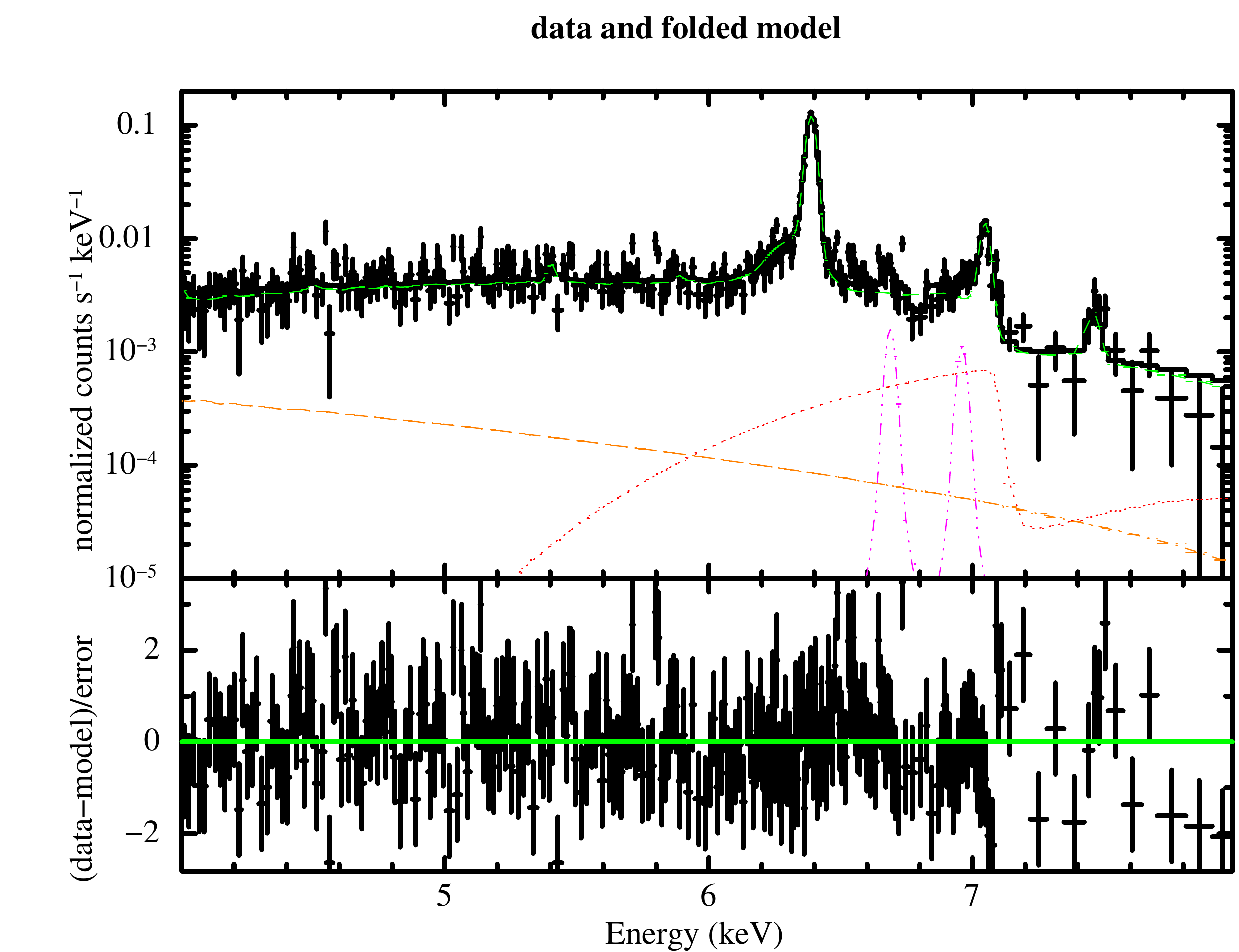}}}
\end{minipage}
\caption{Spectral fitting result using our baseline model. Green-dashed, orange-dashed-dotted, red-dotted, magenta-dashed-dotted-dotted, and black-solid lines represent reflection, scattered, direct, emission line component, and their combination, respectively. The left and right figures show the results obtained by the smooth and clumpy torus structures as the reflection component, respectively.}
\label{fig:result:spec}
\end{center}
\end{figure*}

\begin{figure*}[htbp]
\begin{center}
\begin{minipage}{5cm}
\rotatebox{-90}{\resizebox{4cm}{!}{\includegraphics{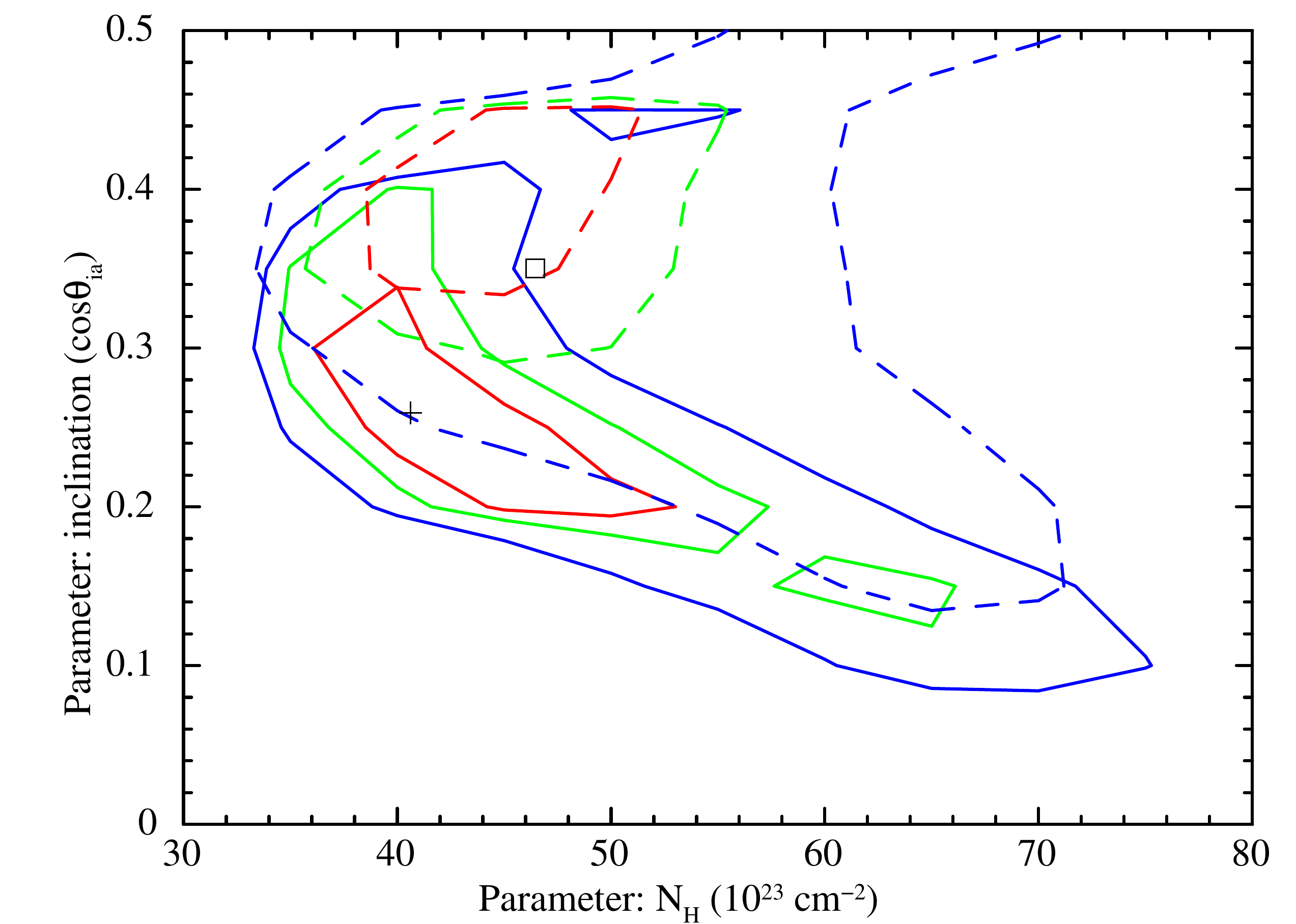}}}
\end{minipage}
\begin{minipage}{5cm}
\rotatebox{-90}{\resizebox{4cm}{!}{\includegraphics{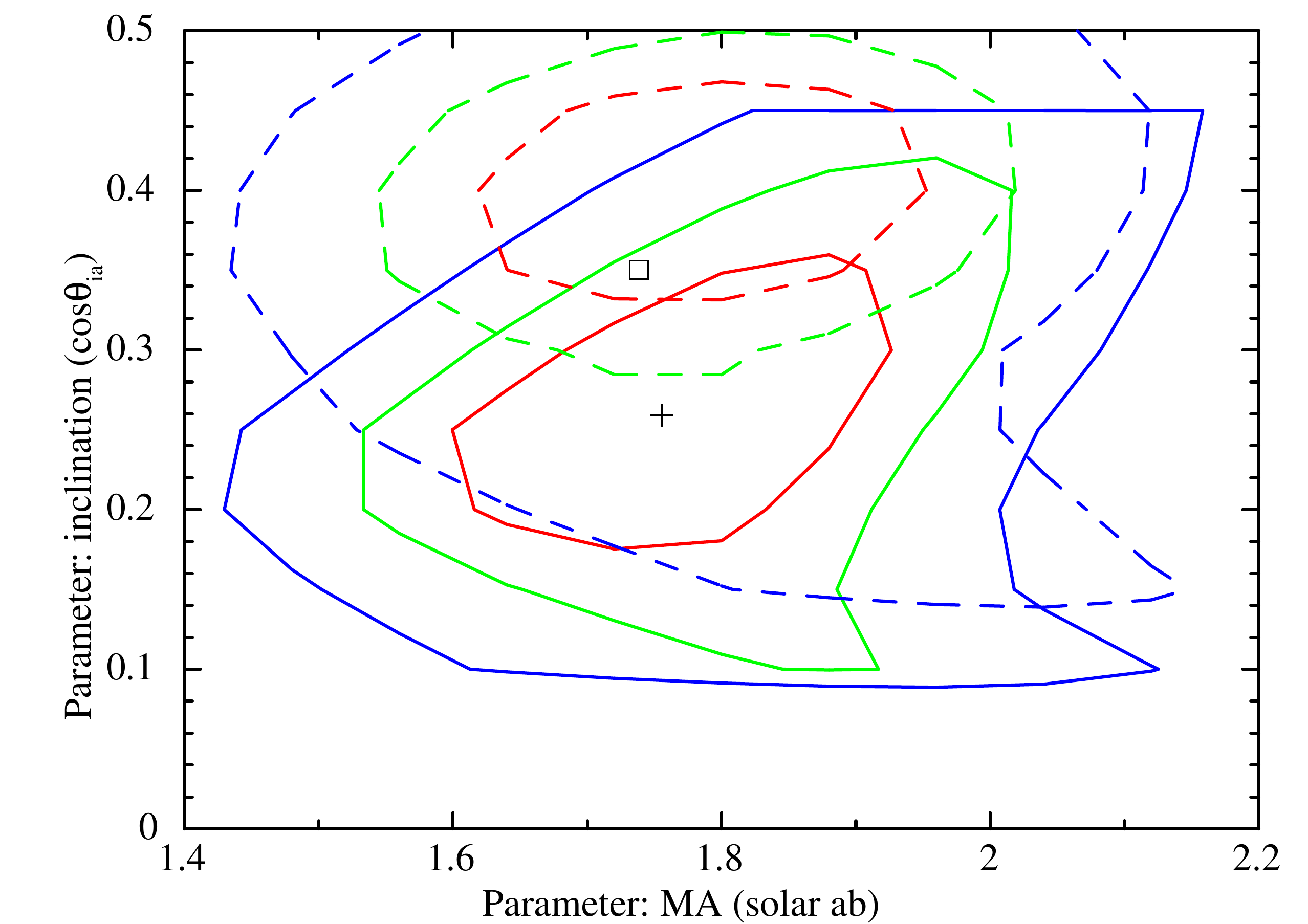}}}
\end{minipage}
\begin{minipage}{5cm}
\rotatebox{-90}{\resizebox{4cm}{!}{\includegraphics{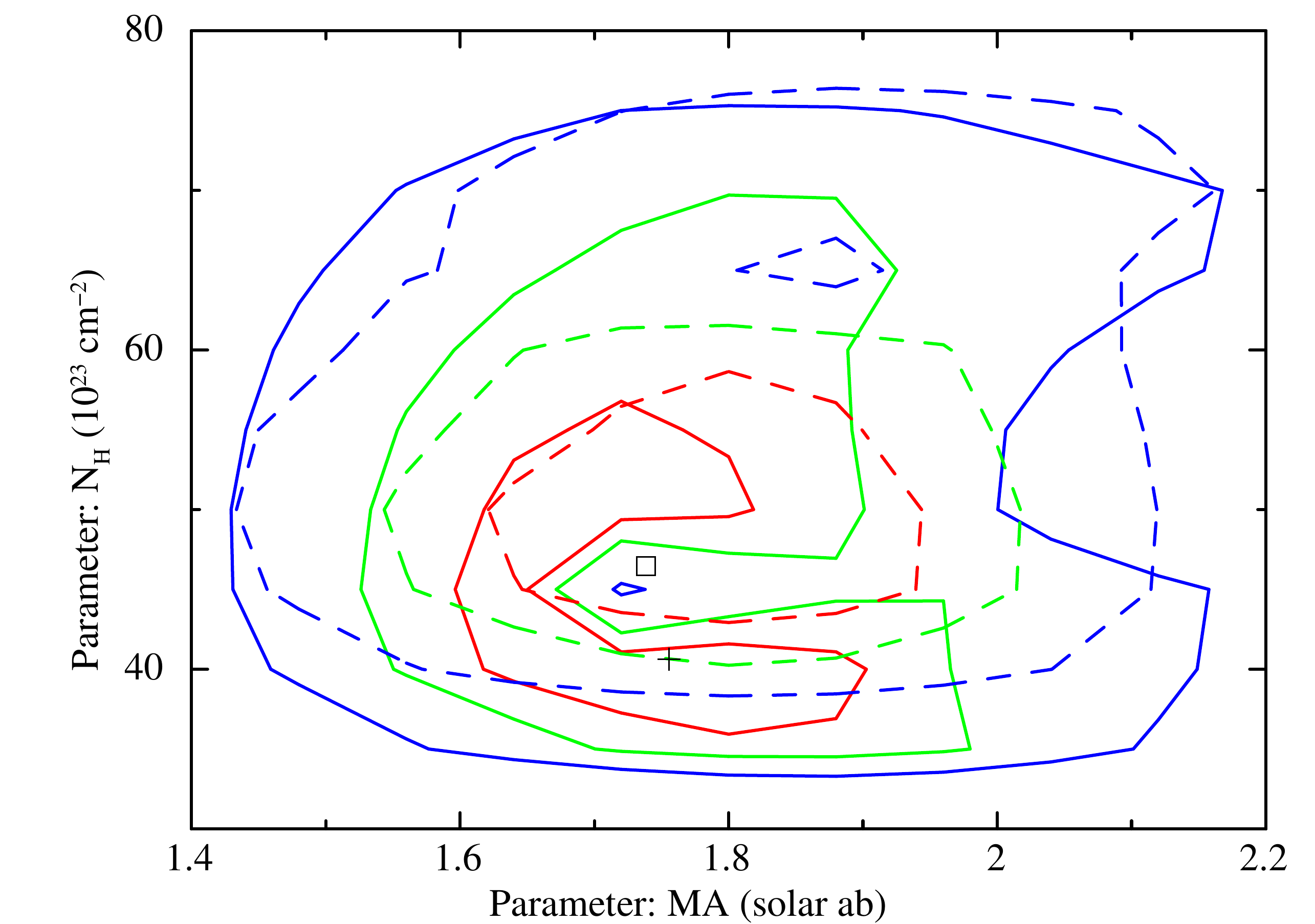}}}
\end{minipage}
\caption{Confidence contours of for column density, inclination angle $\cos \theta_{ia}$, 
and metal abundance. Solid and dashed lines show the result obtained by using the smooth torus and clumpy torus model as the reflection component, respectively. 
Red, green, and blue contours show 68, 90, and 99\% confidence regions, respectively.}
\label{fig:result:contour}
\end{center}
\end{figure*}

Figure \ref{fig:result:spec} shows the {\it Chandra HETG} spectrum and fitting result of our baseline model, together with the 
each spectral component, as we described in \S\ref{sec:analysis}. We examined two torus models, the smooth and clumpy torus.
As shown in the figure \ref{fig:result:spec}, both torus models fit the observed data well and any significant differences did not appear
in this analysis. The summary of the obtained spectral parameters is given in the table \ref{tab:result:fit}. We obtained the column density of
4.1$^{+1.9}_{-0.6}$ and 4.7$^{+1.1}_{-0.5}$  $\times$ 10$^{24}$ cm$^{-2}$ for both torus models and they are consistent within the statistical errors.
These values are consistent with the previous picture that 
this object is in the Compton-thick state and similar to the value that was
reported based on the broadband spectroscopy by {\it Suzaku} (4.6$^{+0.47}_{-0.23}$ $\times 10^{24}$ cm$^2$; \cite{2009ApJ...691..131Y}), and by {\it NuSTAR}  (6-10 $\times 10^{24}$ cm$^2; $\cite{2014ApJ...791...81A}). 
We successfully constrain the inclination angle to be cos$\theta_{ia}$ with the values of 0.26$^{+0.12}_{-0.10}$ and 0.35$^{+0.08}_{-0.08}$ for the smooth and clumpy torus models, respectively, even if we free this parameter. 
Since the torus mid-plane and our line of sight are parallel for cos$\theta_{ia}$ = 0, 
the obtained inclination angle means a nearly
edge-on view of the torus. 
Interestingly, we also constrain the metal abundance for both the smooth and clumpy torus models to 1.75$^{+0.19}_{-0.17}$ and 1.74$^{+0.16}_{-0.16}$ solar, respectively. 
Our baseline model successfully constrains not only the absorption column density but also inclination angle and even metal abundance, despite only analyzing a limited energy band from 4 to 8 keV. 
As we discussed above, the Compton shoulder diagnostics with our reflection model gives this great constraint.
We try to find any differences between our smooth 
and clumpy torus models to discuss whether Compton shoulder diagnostics is also useful in constraining 
the intra-torus structure. 
Figure \ref{fig:result:contour} shows the comparison of the confidence contours between the smooth and clumpy torus models for these three parameters. We cannot see any clear differences between the smooth and clumpy torus models 
(other than a trend that the clumpy model tends to show a slightly larger $\cos \theta_{ia}$)
even in these parameter spaces and we cannot conclude by our analysis which torus model
is better. \cite{2016ApJ...818..164F} suggested that different behaviors can be observed based on the simulated spectra of the {\it Hitomi SXS}. Therefore, higher-resolution spectroscopy with a much better signal-to-noise ratio will show the difference of such parameter correlations with respect to not only column density but also metal abundance, and it could be very useful in constraining the torus structure.


\begin{table*}[htbp]
\begin{center}
\caption{Summary of the obtained spectral parameter in our analysis. The case of scale factor of the scattered component of 0.001 is applied as the best-fit model and denoted by an asterisk sign in this table.}
\label{tab:result:fit}
\begin{tabularx}{18cm}{lcccccccccl}
\hline
\hline
\multicolumn{11}{l}{\it smooth torus model}\\
&\multicolumn{4}{l}{reflection component}&\multicolumn{3}{l}{direct component}&\multicolumn{2}{l}{scattered component}&\\
& $N_{\rm H}$ [10$^{22}$ cm$^{-2}$] & incl. (cos$\theta_{ia}$) & MA$^a$ &\hspace{0.2cm} & $N_{\rm H}$ [10$^{22}$ cm$^{-2}$] $^b$& $\Gamma$ & \hspace{0.2cm} &$\Gamma$ & scale factor$^c$ &$ \chi^2_\nu$(d.o.f) \\
\hline
&355$^{+50.0}_{-36.8}$ & 0.24$^{+0.17}_{-0.08}$ & 1.98$^{+0.20}_{-0.19}$ & & 310 & 1.9 (fixed) & &1.9 (fixed) & 0.003 & 1.47 (580)\\
$\ast$&406$^{+187}_{-56.7}$ & 0.26$^{+0.12}_{-0.10}$ & 1.75$^{+0.19}_{-0.17}$ & & 348 &                  & &                  & 0.001 & 1.43 (580)\\
&410$^{+188}_{-53.2}$ & 0.27$^{+0.12}_{-0.09}$ & 1.71$^{+0.21}_{-0.19}$ & & 340 &                  & &                  & 0.0003 & 1.41 (580)\\
&411$^{+184}_{-52.4}$ & 0.28$^{+0.11}_{-0.09}$ & 1.70$^{+0.21}_{-0.19}$ & & 339 &                  & &                  & 0.0002 & 1.43 (580)\\
\hline 
\multicolumn{11}{l}{\it clumpy torus model}\\
&\multicolumn{4}{l}{reflection component}&\multicolumn{3}{l}{direct component}&\multicolumn{2}{l}{scattered component}&\\
& $N_{\rm H}$ [10$^{22}$ cm$^{-2}$] & incl. (cos$\theta_{ia}$) & MA$^a$ &\hspace{0.2cm} & $N_{\rm H}$ [10$^{22}$ cm$^{-2}$] $^b$& $\Gamma$ & \hspace{0.2cm} &$\Gamma$ & scale factor$^c$ &$ \chi^2_\nu$(d.o.f) \\
\hline
&404$^{+40.7}_{-32.4}$ & 0.34$^{+0.04}_{-0.09}$ & 1.98$^{+0.17}_{-0.18}$ & & 296 & 1.9 (fixed) & &1.9 (fixed) & 0.003 & 1.46 (580)\\
$\ast$&465$^{+114}_{-54.3}$ & 0.35$^{+0.08}_{-0.08}$ & 1.74$^{+0.16}_{-0.16}$ & & 326 &                  & &                  & 0.001 & 1.44 (580)\\
&494$^{+105}_{-61.8}$ & 0.36$^{+0.08}_{-0.07}$ & 1.66$^{+0.16}_{-0.16}$ & & 338 &                  & &                  & 0.0003 & 1.43 (580)\\
&499$^{+113}_{-63.3}$ & 0.37$^{+0.07}_{-0.07}$ & 1.65$^{+0.15}_{-0.16}$ & & 338 &                  & &                  & 0.0002 & 1.44 (580)\\
\hline
\multicolumn{11}{l}{\scriptsize a: MA means the metal abundance which is scaled to the solar value}\\
\multicolumn{11}{l}{\scriptsize b: the column density for the direct component is related to that of the reflection component as a function of $N_{\rm H, direct} = N_{\rm H, refl} \times \sqrt{1-4cos\theta_{ia}}$,}\\
\multicolumn{11}{l}{~~~~\scriptsize see the text for details.}\\
\multicolumn{11}{l}{\scriptsize c: normalization of the scattered component is scaled from that of the direct component by this factor.}\\
\end{tabularx}
\end{center}
\end{table*}

\section{Discussion} \label{sec:discussion}

Since several torus parameters such as the inclination angle, metal abundance and column density 
are degenerate, many X-ray studies for modeling the torus structure have assumed a fixed parameter sets, e.g., solar abundance 
(\cite{2006A&A...455..153M,2014ApJ...791...81A}).
We have successfully constrained these three parameters without any restrictions utilizing the combination of the {\it Chandra HETG} high-resolution spectroscopic data and spectral modeling of the Compton shoulder based on the detailed Monte Carlo simulation even when we used data only 
around the Fe-K$_\alpha$ emission line.  This means that the Compton shoulder could be a strong diagnostics tool for analyzing the torus structure. 
\cite{2016ApJ...818..164F} have already suggested 
this possibility by showing the 
dependence of the structure of the Compton shoulder and other parameters 
based on the detailed studies of the simulated reflection spectral model. 
For example, the ratio of the intensity between the Fe-K$_\alpha$ line core and the Compton shoulder 
depends on the metal abundance in a 
manner different from the $N_{\rm H}$ dependency of the Fe-K$_\alpha$ flux.
In this study, we have proved this concept by applying this reflection model to the actual observational data. Our analysis showed that the metal abundance of the Circinus galaxy is slightly higher than that of the solar abundance  (e.g., 1.75$^{+0.19}_{-0.17}$  for the smooth torus case). 
This over-solar abundance we obtained through this analysis is 
consistent with the metallicity in narrow-line regions (over 100pc$-$1kpc scales in general) 
of other AGNs [(1-4) solar; \cite{2006MNRAS.371.1559G,2014MNRAS.438.2828D}]. 
In this study, we estimate the metallicity of the torus, on much smaller scale (tens of parsecs).

Star formation in clumpy tori may play a major role in angular momentum 
transport in tori (e.g., \cite{2003A&A...404...21V,2008A&A...491..441V}). 
\cite{2013A&A...560A..34W} 
developed such a model to explain nearby AGNs, obtaining 
a star formation rate of $\sim$\,1\,$M_\odot$\,yr$^{-1}$ in tori.
Our measurement implies that the star formation in the torus should not be 
at an enormous rate, if the star formation governs the gas inflow within the torus.


Another important challenge of this study is the investigation of the detailed torus structure by applying the different torus modeling, the smooth and clumpy models, to the observational data. Unfortunately, we cannot obtain a significant discrimination of these two torus models with this analysis. 
Broadband spectroscopy up to the hard X-ray energy band to detect the direct component could bring a further constraint on the column density of the torus independently. Therefore, the simultaneous combination of the high-resolution spectroscopy around the Fe-K$_\alpha$ line and the hard X-ray observation is important. It is encouraging to perform a deep {\it Chandra HETG} observations for the type II Seyfert galaxy and the simultaneous observations by {\it NuSTAR} and it is also interesting to compare the parameters that are determined by such broad-band continuum and our Compton shoulder diagnostics, or further high-resolution spectroscopy by the future missions of {\it XARM} (\cite{Tashiro+18SPIE}) and {\it ATHENA} could provide a finer spectral modeling of both the Compton shoulder and the continuum.\\



\section{Conclusion} \label{sec:conclusion}

In this study, we have demonstrated that various torus parameters, including 
the hydrogen column density at the mid-plane, $N_{\rm H}$, the inclination angle, and metal abundance, can be constrained independently with the {\it Chandra HETG} high-resolution spectroscopy of the Compton shoulder using only the limited energy band around the Fe-K$_\alpha$ emission line. We applied the developed Monte-Carlo-based X-ray reflection spectral model to the Circinus galaxy, and obtained the hydrogen column density of the molecular torus as  $\sim$ 4 $\times$ 10$^{24}$ cm$^{-2}$, which is consistent with that reported based on the broadband spectroscopy. In addition, the inclination angle and the metal abundance were successfully determined without fixing other parameters. The obtained inclination angle 
means that this object is viewed at nearly an edge-on geometry. The important result of this study is that we constrained the metal abundance of the molecular torus to be slightly over-solar abundance ($\sim$ 1.75 solar), which could give information about the evolution of an environment around the super-massive black hole, such as the star formation history around the molecular torus. Further higher-resolution and spatially resolved spectroscopy, which would be brought by future missions such as {\it XARM} and {\it ATHENA}, can measure not only these parameters from many other type II Seyferts, but also can give a discrimination of the detailed torus geometry such as smooth and clumpy density profiles.


\bibliography{mybibfile_v20180622.bib} 

\begin{thebibliography}{}
\expandafter\ifx\csname natexlab\endcsname\relax\def\natexlab#1{#1}\fi
\providecommand{\url}[1]{\href{#1}{#1}}
\providecommand{\dodoi}[1]{doi:~\href{http://doi.org/#1}{\nolinkurl{#1}}}
\providecommand{\doeprint}[1]{\href{http://ascl.net/#1}{\nolinkurl{http://ascl.net/#1}}}
\providecommand{\doarXiv}[1]{\href{https://arxiv.org/abs/#1}{\nolinkurl{https://arxiv.org/abs/#1}}}

\bibitem[{{Anders} \& {Grevesse}(1989)}]{1989GeCoA..53..197A}
{Anders}, E., \& {Grevesse}, N. 1989, \gca, 53, 197,
  \dodoi{10.1016/0016-7037(89)90286-X}

\bibitem[{{Antonucci}(1993)}]{1993ARA&A..31..473A}
{Antonucci}, R. 1993, \araa, 31, 473,
  \dodoi{10.1146/annurev.aa.31.090193.002353}

\bibitem[{{Antonucci} \& {Miller}(1985)}]{1985ApJ...297..621A}
{Antonucci}, R.~R.~J., \& {Miller}, J.~S. 1985, \apj, 297, 621,
  \dodoi{10.1086/163559}

\bibitem[{{Ar{\'e}valo} {et~al.}(2014){Ar{\'e}valo}, {Bauer}, {Puccetti},
  {Walton}, {Koss}, {Boggs}, {Brandt}, {Brightman}, {Christensen}, {Comastri},
  {Craig}, {Fuerst}, {Gandhi}, {Grefenstette}, {Hailey}, {Harrison}, {Luo},
  {Madejski}, {Madsen}, {Marinucci}, {Matt}, {Saez}, {Stern}, {Stuhlinger},
  {Treister}, {Urry}, \& {Zhang}}]{2014ApJ...791...81A}
{Ar{\'e}valo}, P., {Bauer}, F.~E., {Puccetti}, S., {et~al.} 2014, \apj, 791,
  81, \dodoi{10.1088/0004-637X/791/2/81}

\bibitem[{{Arnaud}(1996)}]{1996ASPC..101...17A}
{Arnaud}, K.~A. 1996, in Astronomical Society of the Pacific Conference Series,
  Vol. 101, Astronomical Data Analysis Software and Systems V, ed. G.~H.
  {Jacoby} \& J.~{Barnes}, 17

\bibitem[{{Awaki} {et~al.}(2000){Awaki}, {Ueno}, {Taniguchi}, \&
  {Weaver}}]{2000ApJ...542..175A}
{Awaki}, H., {Ueno}, S., {Taniguchi}, Y., \& {Weaver}, K.~A. 2000, \apj, 542,
  175, \dodoi{10.1086/309516}

\bibitem[{{Awaki} {et~al.}(2008){Awaki}, {Anabuki}, {Fukazawa}, {Gallo},
  {Ikeda}, {Isobe}, {Itoh}, {Kunieda}, {Makishima}, {Markowitz}, {Miniutti},
  {Mizuno}, {Okajima}, {Ptak}, {Reeves}, {Takahashi}, {Terashima}, \&
  {Yaqoob}}]{2008PASJ...60S.293A}
{Awaki}, H., {Anabuki}, N., {Fukazawa}, Y., {et~al.} 2008, \pasj, 60, S293,
  \dodoi{10.1093/pasj/60.sp1.S293}

\bibitem[{{Bianchi} {et~al.}(2006){Bianchi}, {Guainazzi}, \&
  {Chiaberge}}]{2006A&A...448..499B}
{Bianchi}, S., {Guainazzi}, M., \& {Chiaberge}, M. 2006, \aap, 448, 499,
  \dodoi{10.1051/0004-6361:20054091}

\bibitem[{{Bianchi} {et~al.}(2002){Bianchi}, {Matt}, {Fiore}, {Fabian},
  {Iwasawa}, \& {Nicastro}}]{2002A&A...396..793B}
{Bianchi}, S., {Matt}, G., {Fiore}, F., {et~al.} 2002, \aap, 396, 793,
  \dodoi{10.1051/0004-6361:20021414}

\bibitem[{{Bland-Hawthorn} {et~al.}(1991){Bland-Hawthorn}, {Sokolowski}, \&
  {Cecil}}]{1991ApJ...375...78B}
{Bland-Hawthorn}, J., {Sokolowski}, J., \& {Cecil}, G. 1991, \apj, 375, 78,
  \dodoi{10.1086/170170}

\bibitem[{{Bland-Hawthorn} \& {Voit}(1993)}]{1993RMxAA..27...73B}
{Bland-Hawthorn}, J., \& {Voit}, G.~M. 1993, \rmxaa, 27, 73

\bibitem[{{Du} {et~al.}(2014){Du}, {Wang}, {Hu}, {Valls-Gabaud}, {Baldwin},
  {Ge}, \& {Xue}}]{2014MNRAS.438.2828D}
{Du}, P., {Wang}, J.-M., {Hu}, C., {et~al.} 2014, \mnras, 438, 2828,
  \dodoi{10.1093/mnras/stt2386}

\bibitem[{{Furui} {et~al.}(2016){Furui}, {Fukazawa}, {Odaka}, {Kawaguchi},
  {Ohno}, \& {Hayashi}}]{2016ApJ...818..164F}
{Furui}, S., {Fukazawa}, Y., {Odaka}, H., {et~al.} 2016, \apj, 818, 164,
  \dodoi{10.3847/0004-637X/818/2/164}

\bibitem[{{Glass}(2004)}]{2004MNRAS.350.1049G}
{Glass}, I.~S. 2004, \mnras, 350, 1049,
  \dodoi{10.1111/j.1365-2966.2004.07712.x}

\bibitem[{{Groves} {et~al.}(2006){Groves}, {Heckman}, \&
  {Kauffmann}}]{2006MNRAS.371.1559G}
{Groves}, B.~A., {Heckman}, T.~M., \& {Kauffmann}, G. 2006, \mnras, 371, 1559,
  \dodoi{10.1111/j.1365-2966.2006.10812.x}

\bibitem[{{H{\"o}nig} {et~al.}(2006){H{\"o}nig}, {Beckert}, {Ohnaka}, \&
  {Weigelt}}]{2006A&A...452..459H}
{H{\"o}nig}, S.~F., {Beckert}, T., {Ohnaka}, K., \& {Weigelt}, G. 2006, \aap,
  452, 459, \dodoi{10.1051/0004-6361:20054622}

\bibitem[{{Iwasawa} {et~al.}(1997){Iwasawa}, {Fabian}, \&
  {Matt}}]{1997MNRAS.289..443I}
{Iwasawa}, K., {Fabian}, A.~C., \& {Matt}, G. 1997, \mnras, 289, 443,
  \dodoi{10.1093/mnras/289.2.443}

\bibitem[{{Kaspi} {et~al.}(2002){Kaspi}, {Brandt}, {George}, {Netzer},
  {Crenshaw}, {Gabel}, {Hamann}, {Kaiser}, {Koratkar}, {Kraemer}, {Kriss},
  {Mathur}, {Mushotzky}, {Nandra}, {Peterson}, {Shields}, {Turner}, \&
  {Zheng}}]{2002ApJ...574..643K}
{Kaspi}, S., {Brandt}, W.~N., {George}, I.~M., {et~al.} 2002, \apj, 574, 643,
  \dodoi{10.1086/341113}

\bibitem[{{Kawaguchi} \& {Mori}(2010)}]{2010ApJ...724L.183K}
{Kawaguchi}, T., \& {Mori}, M. 2010, \apjl, 724, L183,
  \dodoi{10.1088/2041-8205/724/2/L183}

\bibitem[{{Krolik} \& {Begelman}(1988)}]{1988ApJ...329..702K}
{Krolik}, J.~H., \& {Begelman}, M.~C. 1988, \apj, 329, 702,
  \dodoi{10.1086/166414}

\bibitem[{{Lawrence}(1991)}]{1991MNRAS.252..586L}
{Lawrence}, A. 1991, \mnras, 252, 586, \dodoi{10.1093/mnras/252.4.586}

\bibitem[{{Leipski} {et~al.}(2014){Leipski}, {Meisenheimer}, {Walter}, {Klaas},
  {Dannerbauer}, {De Rosa}, {Fan}, {Haas}, {Krause}, \&
  {Rix}}]{2014ApJ...785..154L}
{Leipski}, C., {Meisenheimer}, K., {Walter}, F., {et~al.} 2014, \apj, 785, 154,
  \dodoi{10.1088/0004-637X/785/2/154}

\bibitem[{{Marinucci} {et~al.}(2013){Marinucci}, {Miniutti}, {Bianchi}, {Matt},
  \& {Risaliti}}]{2013MNRAS.436.2500M}
{Marinucci}, A., {Miniutti}, G., {Bianchi}, S., {Matt}, G., \& {Risaliti}, G.
  2013, \mnras, 436, 2500, \dodoi{10.1093/mnras/stt1759}

\bibitem[{{Massaro} {et~al.}(2006){Massaro}, {Bianchi}, {Matt}, {D'Onofrio}, \&
  {Nicastro}}]{2006A&A...455..153M}
{Massaro}, F., {Bianchi}, S., {Matt}, G., {D'Onofrio}, E., \& {Nicastro}, F.
  2006, \aap, 455, 153, \dodoi{10.1051/0004-6361:20054772}

\bibitem[{{Matsumoto} {et~al.}(2004){Matsumoto}, {Nava}, {Maddox}, {Leighly},
  {Grupe}, {Awaki}, \& {Ueno}}]{2004ApJ...617..930M}
{Matsumoto}, C., {Nava}, A., {Maddox}, L.~A., {et~al.} 2004, \apj, 617, 930,
  \dodoi{10.1086/425566}

\bibitem[{{Matt}(2002)}]{2002MNRAS.337..147M}
{Matt}, G. 2002, \mnras, 337, 147, \dodoi{10.1046/j.1365-8711.2002.05890.x}

\bibitem[{{Matt} {et~al.}(2009){Matt}, {Bianchi}, {Awaki}, {Comastri},
  {Guainazzi}, {Iwasawa}, {Jimenez-Bailon}, \&
  {Nicastro}}]{2009A&A...496..653M}
{Matt}, G., {Bianchi}, S., {Awaki}, H., {et~al.} 2009, \aap, 496, 653,
  \dodoi{10.1051/0004-6361/200811049}

\bibitem[{{Matt} {et~al.}(2004){Matt}, {Bianchi}, {Guainazzi}, \&
  {Molendi}}]{2004A&A...414..155M}
{Matt}, G., {Bianchi}, S., {Guainazzi}, M., \& {Molendi}, S. 2004, \aap, 414,
  155, \dodoi{10.1051/0004-6361:20031635}

\bibitem[{{Molendi} {et~al.}(2003){Molendi}, {Bianchi}, \&
  {Matt}}]{2003MNRAS.343L...1M}
{Molendi}, S., {Bianchi}, S., \& {Matt}, G. 2003, \mnras, 343, L1,
  \dodoi{10.1046/j.1365-8711.2003.06783.x}

\bibitem[{{Mor} \& {Netzer}(2012)}]{2012MNRAS.420..526M}
{Mor}, R., \& {Netzer}, H. 2012, \mnras, 420, 526,
  \dodoi{10.1111/j.1365-2966.2011.20060.x}

\bibitem[{{Nenkova} {et~al.}(2002){Nenkova}, {Ivezi{\'c}}, \&
  {Elitzur}}]{2002ApJ...570L...9N}
{Nenkova}, M., {Ivezi{\'c}}, {\v Z}., \& {Elitzur}, M. 2002, \apjl, 570, L9,
  \dodoi{10.1086/340857}

\bibitem[{{Nenkova} {et~al.}(2008){Nenkova}, {Sirocky}, {Ivezi{\'c}}, \&
  {Elitzur}}]{2008ApJ...685..147N}
{Nenkova}, M., {Sirocky}, M.~M., {Ivezi{\'c}}, {\v Z}., \& {Elitzur}, M. 2008,
  \apj, 685, 147, \dodoi{10.1086/590482}

\bibitem[{{Odaka} {et~al.}(2011){Odaka}, {Aharonian}, {Watanabe}, {Tanaka},
  {Khangulyan}, \& {Takahashi}}]{2011ApJ...740..103O}
{Odaka}, H., {Aharonian}, F., {Watanabe}, S., {et~al.} 2011, \apj, 740, 103,
  \dodoi{10.1088/0004-637X/740/2/103}

\bibitem[{{Odaka} {et~al.}(2016){Odaka}, {Yoneda}, {Takahashi}, \&
  {Fabian}}]{2016MNRAS.462.2366O}
{Odaka}, H., {Yoneda}, H., {Takahashi}, T., \& {Fabian}, A. 2016, \mnras, 462,
  2366, \dodoi{10.1093/mnras/stw1764}

\bibitem[{{Shu} {et~al.}(2011){Shu}, {Yaqoob}, \& {Wang}}]{2011ApJ...738..147S}
{Shu}, X.~W., {Yaqoob}, T., \& {Wang}, J.~X. 2011, \apj, 738, 147,
  \dodoi{10.1088/0004-637X/738/2/147}

\bibitem[{{Suganuma} {et~al.}(2004){Suganuma}, {Yoshii}, {Kobayashi},
  {Minezaki}, {Enya}, {Tomita}, {Aoki}, {Koshida}, \&
  {Peterson}}]{2004ApJ...612L.113S}
{Suganuma}, M., {Yoshii}, Y., {Kobayashi}, Y., {et~al.} 2004, \apjl, 612, L113,
  \dodoi{10.1086/424818}

\bibitem[{Tashiro {et~al.}(2018)Tashiro, Maejima, Toda, \&
  et~al.}]{Tashiro+18SPIE}
Tashiro, M., Maejima, H., Toda, K., \& et~al. 2018, in Space Telescopes and
  Instrumentation 2018: Ultraviolet to Gamma Ray, \procspie

\bibitem[{{Telesco} {et~al.}(1984){Telesco}, {Becklin}, {Wynn-Williams}, \&
  {Harper}}]{1984ApJ...282..427T}
{Telesco}, C.~M., {Becklin}, E.~E., {Wynn-Williams}, C.~G., \& {Harper}, D.~A.
  1984, \apj, 282, 427, \dodoi{10.1086/162220}

\bibitem[{{Torrej{\'o}n} {et~al.}(2010){Torrej{\'o}n}, {Schulz}, {Nowak}, \&
  {Kallman}}]{2010ApJ...715..947T}
{Torrej{\'o}n}, J.~M., {Schulz}, N.~S., {Nowak}, M.~A., \& {Kallman}, T.~R.
  2010, \apj, 715, 947, \dodoi{10.1088/0004-637X/715/2/947}

\bibitem[{{Vollmer} \& {Beckert}(2003)}]{2003A&A...404...21V}
{Vollmer}, B., \& {Beckert}, T. 2003, \aap, 404, 21,
  \dodoi{10.1051/0004-6361:20030436}

\bibitem[{{Vollmer} {et~al.}(2008){Vollmer}, {Beckert}, \&
  {Davies}}]{2008A&A...491..441V}
{Vollmer}, B., {Beckert}, T., \& {Davies}, R.~I. 2008, \aap, 491, 441,
  \dodoi{10.1051/0004-6361:200810446}

\bibitem[{{Watanabe} {et~al.}(2003){Watanabe}, {Sako}, {Ishida}, {Ishisaki},
  {Kahn}, {Kohmura}, {Morita}, {Nagase}, {Paerels}, \&
  {Takahashi}}]{2003ApJ...597L..37W}
{Watanabe}, S., {Sako}, M., {Ishida}, M., {et~al.} 2003, \apjl, 597, L37,
  \dodoi{10.1086/379735}

\bibitem[{{Wutschik} {et~al.}(2013){Wutschik}, {Schleicher}, \&
  {Palmer}}]{2013A&A...560A..34W}
{Wutschik}, S., {Schleicher}, D.~R.~G., \& {Palmer}, T.~S. 2013, \aap, 560,
  A34, \dodoi{10.1051/0004-6361/201321895}

\bibitem[{{Yang} {et~al.}(2009){Yang}, {Wilson}, {Matt}, {Terashima}, \&
  {Greenhill}}]{2009ApJ...691..131Y}
{Yang}, Y., {Wilson}, A.~S., {Matt}, G., {Terashima}, Y., \& {Greenhill}, L.~J.
  2009, \apj, 691, 131, \dodoi{10.1088/0004-637X/691/1/131}

\bibitem[{{Yaqoob} {et~al.}(2015){Yaqoob}, {Tatum}, {Scholtes}, {Gottlieb}, \&
  {Turner}}]{2015MNRAS.454..973Y}
{Yaqoob}, T., {Tatum}, M.~M., {Scholtes}, A., {Gottlieb}, A., \& {Turner},
  T.~J. 2015, \mnras, 454, 973, \dodoi{10.1093/mnras/stv2021}

\end{thebibliography}
\bibliographystyle{aasjournal}
\end{document}